# SCALED QUANTIZATION FOR THE VISION TRANSFORMER

Yangyang Chang[1] and Gerald E. Sobelman[2]

[1,2]Department of Electrical Engineering and Computer Science, University of Minnesota, Minneapolis, MN, USA
[1]chan1729@umn.edu
[2]sobelman@umn.edu

## ABSTRACT

*Quantization using a small number of bits shows promise for reducing latency and memory usage in deep neural networks. However, most quantization methods cannot readily handle complicated functions such as exponential and square root, and prior approaches involve complex training processes that must interact with floating-point values. This paper proposes a robust method for the full integer quantization of vision transformer networks without requiring any intermediate floating-point computations. The quantization techniques can be applied in various hardware or software implementations, including processor/memory architectures and FPGAs.*

## 1. INTRODUCTION

Computer vision applications have advanced using deep convolutional neural networks (CNNs) [1, 2, 22]. Many types of CNN structures with various convolution operators (e.g., depth-wise separable and point-wise convolution operators) have been developed [3-4]. Recently, inspired by the use of transformers in natural language processing (NLP) [5], the vision transformer paradigm has achieved performance similar to or better than that of CNNs [6-9].

Quantization is a technique in which some or all of the calculations are done using fixed-point operands (e.g., INT8 or INT16). These require fewer complex implementations than for floating-point calculations (e.g., FP32 or FP64). In the context of neural network architectures, there are two major quantization processes: post-training quantization [10,11,18] and quantization-aware training [12,13,17]. The method of post-training quantization is similar to a model compression technique without fine-tuning. Post-training quantization has the ability to reduce latency, power and model size with only a limited degradation in accuracy [14]. In this technique, a model is trained using floating-point values and it is then quantized. Quantization-aware training, on the other hand, is a training method in which quantization is used directly during the design process [12, 13, 17].

Refs. [15,16,17,18] have applied various quantization techniques to the transformer. For example, [16] uses binary code weights and a full precision scale to enable efficient matrix multiplication, while [17] applies a fixed-point retraining process to obtain the final quantized architecture. Ref. [18] focuses on the post-training quantization of the vision transformer by measuring the similarity between output feature maps in the linear operation, searching optimal quantization intervals in the self-attention, and correcting the bias. Although the final quantized architectures of [17, 18] can achieve a smaller memory size, intermediate floating-point calculations must be re-quantized after some operations, which consumes additional computational resources. Ref.

**Algorithm 1. The Overflow Method**

---

Input scalar value: $INT_1, INT_2, S_1, S_2, INT_T$
($INT_T$ is a scalar value of temporary storage)
($INT_1, INT_2, S_1, S_2$ have the same size $P$)
1: If $INT_{out}$ is overflow:
2:   Save the operations of $(INT_1, INT_2)$ to $INT_T$
3:   Only keep $P$ MSBs and save to $INT_{out}$
4: EndIf
5: $S_{out}$ = scale operation $(S_1, S_2)$ – number of bits overflowed
Output scalar value: $INT_{out}, S_{out}$

---

[15] uses 8-bit integer inference in the transformer model. However, the memory space requirements are not fully considered. The method of [21] provides a quantization model for CNNs that are implemented on FPGAs, but the proposed scheme is not directly applicable to other complex operators in the vision transformer.

In order to find an efficient quantization method for the vision transformer, this paper proposes a post-training quantization method having the following benefits: 1) It provides a multi-bit quantization process for all operators of the vision transformer without requiring any intermediate floating-point operations. 2) The differences between the quantized and floating-point numerical values for each operator are determined explicitly. 3) It can be efficiently implemented in either dedicated hardware, such as an FPGA, or on a CPU.

## 2. BACKGROUND: VISION TRANSFORMERS

The vision transformer (ViT) without any recurrence or convolutions was proposed in [6]. The ViT only uses the encoder portion of the transformer and it also includes some small changes as compared to the original NLP transformer of [5]. The ViT performs well but it has some limitations. It uses fixed-size coarse image patches (32 by 32 pixels per patch) as input, which leads to reduced resolution [7]. Another implementation, DeiT [9], also uses a fixed patch size. To address these limitations, the Pyramid Vision Transformer (PVT) [7] uses four stages to generate feature maps at different scales, according to a progressive shrinking strategy. This approach enables the PVT to be trained on a coarse image and still achieve a fine output image while reducing the computations associated with large feature maps. In the later design of CoaT [8], fine-to-coarse, coarse-to-fine, and cross-scale attention can all be implemented using a set of serial and parallel blocks.

These transformer networks all use similar operators, including GELU, layer normalization, softmax and linear transformation. Here, we use CoaT [8] as a specific example and show how to quantize these operators for that system.

## 3. QUANTIZATION ALGORITHMS

### 3.1. Overflow

The overflow method is shown in Algorithm 1. $INT_i$ and $S_i$ are the integer portion and scale, respectively, of the two operands in an addition, subtraction, multiplication or division. In case of overflow, only the $P$ MSBs are kept. Thus, it requires additional temporary memory to store the overflow bits. Each result can be kept in temporary register, which can then be re-used to store the next result. Alternatively, all elements of an overflow array can be used to retain the $P$ MSBs of each value.

### 3.2. Basic Operations

**Algorithm 2. ScaleDivison**

Input scalar value: $INT_1, S_1, INT_2, Sum, S_{sum}, T$
1: If Sum is not larger than $2^P - 1$ or T != 5:
2:   If $INT_1 > INT_2$:
3:     Find Q and Rem such that $INT_1 = INT_2$ * Q + Rem
4:     $Sum, S_{sum}$ = ScaleSummation (Q , $S_1$, $Sum, S_{sum}$)
5:     T = T + 1
6:     If Rem != 0:
7:      $Sum, S_{sum}$=ScaleDivison (Rem, $S_1, INT_2, Sum, S_{sum}, T$)
8:     EndIf
9:   ElseIf $INT_1 < INT_2$:
10:    The smallest $INT_1 = INT_1 * 2^{C_T}$ larger than $INT_2$ and
      $S_1 = S_1 + C_T$
11:   $Sum, S_{sum}$= ScaleDivison ($INT_1, S_1, INT_2, Sum, S_{sum}, T$)
12: Else:
13:   $Sum, S_{sum}$= ScaleSummation (1 , $S_1$, $Sum, S_{sum}$)
14: EndIf
15: EndIf
16: Return $Sum, S_{sum}$

Multiplication: Two quantized operands $F_1$ and $F_2$ can be multiplied using (1), which we refer to as *ScaleMultiplication*:

$$F_1 * F_2 = \frac{INT_1}{2^{S_1}} * \frac{INT_2}{2^{S_2}} = \frac{INT_1 * INT_2}{2^{S_1+S_2}} = \frac{INT_{out}}{2^{S_{out}}} \quad (1)$$

Addition/Subtraction: If the larger scale $S_2$ is chosen, the sum or difference is given in (2), which shows both the addition and subtraction computations. We refer to these operations as *ScaleSummation* and *ScaleSubtraction*. We note that $2^x$ is easily achieved by an arithmetic right or left shift operator.

$$INT_{out} = (2^{S_2-S_1} INT_1 + INT_2); S_{out} = S_2$$
$$INT_{out} = (2^{S_2-S_1} INT_1 - INT_2); S_{out} = S_2 \quad (2)$$

Division: The recursive algorithm used, which we refer to as *ScaleDivision*, only accepts inputs having the same sign. If the inputs have different signs, their absolute values will be fed into the algorithm, and the result will be negated. There are 6 inputs: $INT_1, S_1, INT_2, Sum, S_{sum}, T$. $INT_1, INT_2$ are the dividend and divisor, respectively, and $S_1$ is the scale of $INT_1$. $Sum$ and $S_{sum}$ are the quotient value $(INT_1, S_1)/(INT_2, S_2)$ and its scale, respectively. The parameter $T$ controls the resolution of the output. The recursive division procedure is summarized in (3) and implemented as Algorithm 2. Initially, $T, Sum, S_{sum}$ are all 0. $Q$ is the quotient, $Rem$ is the remainder, and $C_1, C_2, \ldots$ are scale values used in each recursive call. If $Rem$ is 0, the program terminates; otherwise, it performs another recursive call and $T$ is increased by 1. When $T$ reaches the specified value or when $Rem$ is 0, the final values of $Sum, S_{sum}$ are output as the quotient results.

$$\begin{aligned}
Sum = &\ Q(INT_1, INT_2) \\
&[+]\ Q(2^{C_1}*Rem(INT_1, INT_2), INT_2) \text{ with the scale } C_1 \\
&[+]\ Q(2^{C_2}*Rem(2*C*Rem(INT_1, INT_2), INT_2), INT_2) \text{ with the scale } C_1+C_2 \\
&[+]\ \ldots
\end{aligned} \quad (3)$$

## 4. MEMORY REDUCTION AND NUMBER OF OPERATIONS

For the examples considered here, $P$ is set to 8. The memory size used to store the scale is 5 bits, allowing scale values from -16 to 15, inclusive. The temporary memory needed to store overflow data is based on the algorithm. If $INT_T$ is reused for all elements, only one temporary memory location of size $2P$ is required. Therefore, the total memory reduction factor compared with FP64 is 64/(8+5) = 4.923.

*ScaleSummation* and *ScaleSubtraction* use both maximum and multiplication operations. The *ScaleMultiplication* requires addition of the scales as an extra operation. The *ScaleDivision* requires the extra addition and subtraction operations. Although the number of operations is increased, their implementations use simple procedures. Therefore, the quantized calculations with shortened bit lengths will be more efficient than using floating-point calculations [14, 19]

## 5. QUANTIZATION FOR THE VISION TRANSFORMER

The specific vision transformer structure considered is CoaT [8]. For each operator, the quantized results are compared with 64-bit floating-point (FP64) calculations. All quantization is performed using the methods described in Section III.

### 5.1. Convolutional Operator and Layer Normalization

In CoaT, the raw input image is sent to the patch embedding layer. This layer is composed of two operations, namely project and layer normalization. The project operation utilizes the 2D convolution operator, which in turn includes element-wise multiplication and partial summation. The layer normalization calculation over a mini-batch of inputs can be summarized according to [23] as in (4).

$$\frac{input - E[input]}{\sqrt{Var[input] + \epsilon}} * \gamma + \beta \qquad (4)$$

Here $\gamma$ and $\beta$ are learnable affine transformation parameters, and $\epsilon$ is included to prevent division by a very small value. The layer normalization requires the calculation of the inverse square root, average, and variance.

Consider the 2D convolution operator with C input channels. The element-wise multiplication and partial summation are done using *ScaleMultiplication*, *ScaleSummation* and *ScaleSubtraction*. Eq. (5) shows the quantized element-wise multiplication of the $(i_{th}, j_{th})$ element of the $xth$ kernel $W_{ij}^x$ and the $(a_{th}, b_{th})$ element $SI_{ab}^y$ of the $yth$ input channel. Here $Eout_{ab}^y$ with the scale $SE_{ab}^y$ is the element-wise output.

$$Eout_{ab}^y = W_{ij}^x * Input_{ab}^y; SE_{ab}^y = SW_{ij}^x + SI_{ab}^y \qquad (5)$$

Then, the quantized partial summation is calculated by the *ScaleSummation* operator. Each input scale must be matched by choosing one maximum scale level $SF_{ab}^y$ from the set $\{SE_0^y, \ldots, SE_{ab}^y, \ldots, SE_{i*j}^y\}$. The output of the $(a_{th}, b_{th})$ element $Fout_{ab}^y$ with the scale $SF_{ab}^y$ from this $xth$ kernel is shown in (6).

$$Fout_{ab}^y = \frac{Eout_0^y}{2^{SE_0^y - SP_{ab}^y}} + \frac{Eout_1^y}{2^{SE_1^y - SP_{ab}^y}} + \cdots + \frac{Eout_{i*j}^y}{2^{SE_{i*j}^y - SP_{ab}^y}} \qquad (6)$$

Using the same concepts, the further two *ScaleSummation* operations are calculated by summing up all the $(a_{th}, b_{th})$ elements from the input channels and then adding the quantized bias. Here, the maximum scale level $SG_{ab}^x$ is from the set $\{SF_{ab}^0, \ldots, SF_{ab}^y, \ldots, SF_{ab}^C\}$; $CS_{ab}^x$ is maximum scale from the set $\{SB_{ab}^x, SG_{ab}^x\}$; $Gout_{ab}^x$ is the *ScaleSummation* output after adding up the all ($a_{th}$, $b_{th}$) elements; and $out_{ab}^x$ is the 2D convolutional output at the ($a_{th}$, $b_{th}$) position of the $x_{th}$ channel.

Table 1. Mean squared error between the quantized operators and the PyTorch operators

| Row | Quantized Operator | MSE |
|---|---|---|
| 1 | 2D Convolution (I = 3, O = 3) | 7.64e-5 |
| 2 | 2D Convolution (I = 3, O = 9) | 6.63e-5 |
| 3 | 2D Convolution (I = 3, O = 1) | 6.89e-5 |
| 4 | Layer Normalization | 0.0015 |
| 5 | 2D Depthwise Convolution (I =3, O = 3) | 2.80e-5 |
| 6 | 2D Depthwise Convolution (I =3, O = 9) | 0.0059 |
| 7 | Linear Transformation (I =3, O = 9) | 3.31e-5 |
| 8 | Linear Transformation (I =3, O = 1) | 1.79e-5 |
| 9 | Linear Transformation (I =3, O = 3) | 3.45e-5 |
| 10 | Softmax | 1.79e-5 |
| 11 | GELU based on (16.1) | 0.0066 |
| 12 | GELU based on (16.2) | 0.0002 |

$$Gout_{ab}^x = \frac{Fout_{ab}^0}{2^{SF_{ab}^0 - SG_{ab}^x}} + \cdots + \frac{Fout_{ab}^y}{2^{SF_{ab}^Y - SG_{ab}^x}} + \cdots + \frac{Fout_{ab}^C}{2^{SF_{ab}^C - SG_{ab}^x}}$$

$$out_{ab}^x = \frac{Gout_{ab}^x}{2^{SG_{ab}^x - CS_{ab}^x}} + \frac{B_{ab}^x}{2^{SB_{ab}^x - CS_{ab}^x}} \quad (7)$$

Each result is checked for overflow using Algorithm 1.

In all of our numerical experiments, 100 images from ImageNet [24] are randomly selected. The sizes of the images are modified to 224x224 by the CenterCrop function in PyTorch prior to processing. Every image is tested 25 times, each time with a new set of random weights and biases. We use a batch size of 1 and a kernel size of 1. $O$ is the size of the output hidden dimension (i.e., number of output channels), $I$ is the size of the input hidden dimension (i.e., number of input channels), $K$ is the kernel size (i.e., length or width), $B$ is the batch size and $C$ is the input image size (i.e., length or width).

The mean squared error (MSE) results for the quantized convolution operator are listed in the first 3 rows of Table 1. (The values given in the other rows of the table will be discussed in later portions of this section.) The formula used to calculate the MSE is given in (8). Here, $Output_{Qij}$ and $Scale_{Qij}$ are the $i_{th}$ element of output tensors at the $j_{th}$ hidden dimension for the quantized convolutional operator; and $Output_{Pij}$ is the $i_{th}$ element of output tensor at the $j_{th}$ hidden dimension obtained with FP64 computations using PyTorch.

$$MSE = \frac{\sum_{j=1}^{j=O} \sum_{i=1}^{i=2C} (\frac{Output_{Qij}}{Scale_{Qij}} - Output_{Pij})^2}{O * 2C} \quad (8)$$

For layer normalization, the quantization of the mean and variance can be implemented using the *ScaleMultiplication*, *ScaleSummation*, *ScaleSubtraction* and *ScaleDivison* operators. The inverse square root is calculated by Newton's method. Letting $y$ be the result of the inverse square root of $x$, we can use Newton's method to find the root using $f(y) = \frac{1}{y^2} - x = 0$. For each iteration

**Algorithm 3. Quantized Newton's Method for the Inverse Square Root**
1: For $x$ iterations:
2: Check overflows of $(Y_j)^3$ and $3Y_j$
3: At the 0th iteration: $(Y_j)^3 * Input_x$
4: At the other iterations: $(Y_j)^3 * Input_x$; Scale = Scale - 3
5: If the result of Line 2 or 3 is overflow:
6:   $Y_{mid}, SY_{mid}$ are modified by Algorithm 1
7: Endif
8: At the 0th iteration: $3Y_j * 2^{2SY_0+SX-SY_{mid}} - Y_{mid}$
9: At the other iterations: $3Y_j * 2^{2SY_0+SX-SY_{mid}-1} - Y_{mid}$
10: If the result of Line 7 or 8 is overflow:
11: $Y, SY$ are modified by Algorithm.1
12: Endif
13: $Y_j = \frac{Y}{2^{2SY_0+SX-SY_{mid}-SY}}$
14: Endfor

Table 2. The quantized Newton's method compared with the floating-point implementation for 1/sqrt(15.25)

| Iteration | Eq. (9) using FP64 | Quantized Approach |
|---|---|---|
| 0 | Y0 = 0.015625<br>X = 15.25 | $Y_0 = 1, SY_0 = 6$<br>$Input_x = 122, SX = 3$ |
| 2 | 0.0350 | $Y_2=5, SY_2=7$ |
| 4 | 0.0772 | $Y_4=12, SY_4=7$ |
| 6 | 0.1576 | $Y_6=25, SY_6=7$ |
| 8 | 0.2426 | $Y_8=33, SY_8=7$ |
| Comparison | 0.2426 vs 0.2561 | 0.2578 vs 0.2561 |

$j$, we can better approximate the next $y_{j+1}$ by (9).

$$y_{j+1} = y_j - \frac{f(y_j)}{f'(y_j)} = y_j - \frac{\frac{1}{y_j^2}-x}{-\frac{2}{y_j^3}} = y_j - \frac{y_j^3 x - y_j}{2} \tag{9}$$

To achieve convergence, the initial approximate $y_0$ must be a small value. In our application, 0.015625 is picked as $y_0$, which is same as $Y_0 = 1$ with $SY_0 = 6$. For any $Input_x$ with its scale $SX$, the quantized calculation from (9) is shown in (10.1). After some rearrangement, it can be expressed more conveniently as (10.2).

$$\frac{Y_{j+1}}{2^{SY_0}} = \frac{3Y_j - \frac{Y_j^3 Input_x}{2^{2SY_0+SX}}}{2*2^{SY_0}} \tag{10.1}$$

$$\frac{Y_{j+1}}{2^{SY_0}} = \frac{3Y_j * 2^{2SY_0+SX} - Y_j^3 Input_x}{2*2^{SY_0}*2^{2SY_0+SX}} \tag{10.2}$$

Here, all of the power-of-2 calculations are implemented by arithmetic shifts. Algorithm 3 gives the precise implementation we have used. Since the power-of-2 calculations may produce large intermediate results, the overflow procedure of Algorithm 1 is invoked to manage those issues.

Table 2 shows an example of the quantized Newton's method compared with floating-point iterations of (9) for the inverse square root of 15.25. The correct answer is 0.2561. Our quantization method gives a closer result after 8 iterations for this example. Similar results showing a faster initial convergence by the quantized algorithm were obtained in a large number of additional numerical experiments.

The fourth row of Table 1 gives the MSE result for layer normalization. Here, $\gamma$ is 1, $\beta$ is 0, and the number of iterations in Newton's method is 20. According to (4), if all quantized inputs are identical, the numerator will be 0. If the numerator is 0, the quantized output is set to $\beta$ without considering the value of $\epsilon$.

### 5.2. Depthwise Convolution Operator

In CoaT, a depthwise convolution, in which each input is convolved with its respective filter, is used in the positional encoding operation. The quantized calculations use the same methods as were described for ordinary convolution except that the number of blocked connections from input channels to output channels is 3. The MSE results are recorded in the fifth and sixth rows of Table 1.

### 5.3. Linear Transformation and Softmax

The major components of a traditional transformer are the stacked layers of self-attention and a feed-forward network. The self-attention is composed of the image tokens (Q, K, V). The formulation for self-attention is shown in (11) [5].

$$(Q, K, V) = Softmax\left(\frac{QK^T}{\sqrt{d_m}}\right) V \qquad (11)$$

In the factorized attention of the CoaT vision transformer [8], the common operators of Softmax, matrix transpose and matrix multiplication are used. Thus, the quantization procedures for these operations can be applied in all similar transformer structures.

In CoaT, the image token is generated and the factorized attention is calculated. The generation of the image tokens is achieved by a linear transformation (e.g., $y = xW^T + b$). The MSE results for the quantized linear transformation are given in rows 7-9 of Table 1.

The calculation of factorized attention is based on $Softmax(K)^T$ and $\frac{Q}{\sqrt{d_m}} \cdot \frac{Q}{\sqrt{d_m}}$, which can be obtained using Algorithms 2 and 3. The Softmax can be implemented by a truncated Maclaurin series. If the first three terms in the series are kept, the Softmax function can be approximated as in (12).

$$\text{Approximated Softmax}(x_i) = \frac{1 + x_i + \frac{x_i^2}{2}}{\sum_j (1 + x_j + \frac{x_j^2}{2})} \qquad (12)$$

Using *ScaleMultiplication* and *ScaleSummation*, the quantized Softmax function is given in (13) with the input integer $X_i$ and its scale $SX_i$. Here, the denominator uses the maximum scale $SX_{max}$ from $\{SX_0, SX_1, \ldots, SX_j\}$ of all inputs $X$ and the scale $SX_i$ is used in the numerator so that the final scale of each input $X_i$ is $\frac{SX_i}{SX_{max}}$.

$$X_i = \frac{2^{2SX_i} + X_i 2^{SX_i} + \frac{X_i^2}{2}}{\sum_j (2^{SX_{max}} + X_j 2^{SX_{max}-SX_j} + 2^{2SX_{max}-SX_j-1})} \qquad (13)$$

In the implementation, the summation in the nominator is calculated first. After checking for overflow, it will be stored in memory. The size of this memory is based on how many of inputs are to be normalized. For example, if the input hidden dimension size is 10, the size of this memory is 10 by P. Then, only one double-size temporary memory is used to add each summation from the first step. Finally, each element of the numerator memory is divided by the temporary memory using ScaleDivison. Three overflow checks are performed. The first is to check for overflow of $X_i^2$. The second is to check for overflow of the summation in the denominator of (13). The third is to check after adding each summation in the numerator. The MSE results for the quantized Softmax operator is given in row 10 of Table 1.

## 5.4. Multlayer Perceptron (Feed-forward network)

A basic feed-forward network can be viewed as a ReLU activation between two linear transformations. The quantized implementation of ReLU is straightforward. Here, we instead focus on the GELU activation that is used in the vision transformer. The GELU can be approximated as in (14) [20].

$$0.5x(1 + \tanh[\sqrt{\frac{2}{\pi}}(x + 0.044715x^3)]) \tag{14}$$

The hyperbolic tangent function can be simplified via the estimation of the exponential function by a Maclaurin series with four terms, as shown in (15).

$$\tanh(x) = \frac{e^x - e^{-x}}{e^x + e^{-x}} = 1 - \frac{2}{e^{2x}+1} = x - \frac{x^3}{3} \tag{15}$$

Using (15), (14) can be simplified to (16.1) or (16.2). Here, $A = \frac{102x}{2^7} + \frac{18x^3}{2^9}$. We can use the *ScaleMultiplication* and *ScaleSummation* operators to perform these calculations.

$$(15) = 0.5x(1 + tanh(0.797884x + 0.035677x^3))$$

$$\approx 0.5x(1 + A + A^3) \tag{16.1}$$

$$\approx 0.5x(1 + A) \tag{16.2}$$

The MSE results for the quantized GELU operator with these two approximations are given in rows 11 and 12 of Table 1.

## 6. CONCLUSIONS

We have presented quantization methods for all of the basic operators in the vision transformer using techniques that do not require any intermediate floating-point calculations or associated de-quantization processes. For each operator, the total memory reduction factor compared with FP64 is 4.923. Although the number of operations is increased due to calculating the scale, our designs are very hardware-friendly due to the use of simple MUX and shift operators. The very small MSE for each operator indicates that our quantization method can be used in place of the floating-point computations in many situations. This will enable efficient implementations of vision transformers to be developed and deployed in new areas of application. In the future work, these proposed quantized operators will be integrated to test the classification accuracy of the entire transformer. Using the optimization technique [25], the memory can be further reduced by optimizing the bit-length of each parameter.

# REFERENCES

[1] K. Fukushima, "Neocognitron: A SELF-ORGANIZING neural network model for a mechanism of pattern recognition unaffected by shift in position," Biological Cybernetics, vol. 36, no. 4, pp. 193–202, 1980.

[2] Y. LeCun et al., "Object recognition with gradient-based learning," Shape, Contour and Grouping in Computer Vision, pp. 319–345, 1999.

[3] A.Howard et al., "MobileNets: Efficient Convolutional Neural Networks for Mobile Vision Applications," arXiv:1704.04861v1 [cs.CV], Apr. 2017.

[4] F.Chollet, "Xception: Deep Learning with Depthwise Separable Convolutions," arXiv:1610.02357v3 [cs.CV], Apr. 2017.

[5] A. Vaswani et al., "Attention Is All You Need," arXiv:1706.03762v5 [cs.CL], Dec. 2017..

[6] A. Dosovitskiy et al., "An Image is Worth 16x16 Words: Transformers for Image Recognition at Scale," arXiv:2010.11929v2 [cs.CV], Jun. 2021.

[7] W.Wang et al., "Pyramid Vision Transformer: A Versatile Backbone for Dense Prediction without Convolutions," arXiv:2102.12122v2 [cs.CV], Aug. 2021.

[8] W. Xu et al., "Co-Scale Conv-Attentional Image Transformers," arXiv:2104.06399v2 [cs.CV], Aug. 2021.

[9] H. Touvron., "Training data-efficient image transformers & distillation through attention," arXiv:2012.12877v2 [cs.CV], Jan. 2021

[10] Z. Liu et al., "Post-Training Quantization for Vision Transformer," arXiv:2106.14156v1 [cs.CV], Jun. 2021.

[11] W. Sung, S.Shin, and K. Hwang, "Resiliency of Deep Neural Networks under Quantization," arXiv:1511.06488v3 [cs.LG], Jan. 2016.

[12] O. Zafrir et al., "Q8BERT: Quantized 8Bit BERT," arXiv:1910.06188v2 [cs.CL], Oct. 2019.

[13] S. Shen et al., "Q-BERT: Hessian Based Ultra Low Precision Quantization of BERT," arXiv:1909.05840v2 [cs.CL], Sep. 2019.

[14] M. Abadi et al., ''TensorFlow: A system for large-scale machine learning,'' in Proc. OSDI, vol. 16, 2016, pp. 265–283.

[15] Y. Lin, et al, "Towards fully 8-bit integer inference for the Transformer model," Proceedings of the Twenty-Ninth International Joint Conference on Artificial Intelligence, 2020.

[16] I. Chung, B. Kim, Y. Choi, S. J. Kwon, Y. Jeon, B. Park, S. Kim, and D. Lee, "Extremely low bit transformer quantization for on-device neural machine translation," Findings of the Association for Computational Linguistics: EMNLP 2020, 2020.

[17] Y. Boo and W. Sung, "Fixed-point optimization of Transformer Neural Network," ICASSP 2020 - 2020 IEEE International Conference on Acoustics, Speech and Signal Processing (ICASSP), 2020.

[18] Z. Liu et al., "Post-Training Quantization for Vision Transformer," arXiv:2106.14156v1 [cs.CV], Jun. 2021.

[19] A. Fog, "Instruction tables: Lists of instruction latencies, throughputs and micro-operation breakdowns for Intel, AMD and VIA CPUs," Debugging & verifying programs -- CS 340D -- Agner Fog's x86 reference material. [Online]. Available: https://www.cs.utexas.edu/users/hunt/class/2018-spring/cs340d/documents/Agner-Fog/.

[20] D. Hendrycks et al., "Gaussian Error Linear Units (GELUs)," arXiv:1606.08415v4 [cs.LG], Jul. 2020.

[21]     Y. Chang and G. E. Sobelman "Genetic Architecture Search for Binarized Neural Networks," IEEE 13th International Conference on ASIC, October, 2019.

[22]     Y. Chang and G. E. Sobelman, "Lightweight CNN Frameworks and their Optimization using Evolutionary Algorithms," International Electrical Engineering Congress, March, 2022.

[23]     A Paszke et al., "PyTorch: An Imperative Style, High-Performance Deep Learning Library," arXiv:1912.01703 [cs.LG] Dec. 2019.

[24]     J. Deng, W. Dong, R. Socher, L.-J., K. Li and L. Fei-Fei, "Imagenet: A large-scale hierarchical image database," IEEE Conference on Computer Vision and Pattern Recognition, pp. 248–255, 2009.

[25]     Y. Chang and G. E. Sobelman, "The class algorithm: Evolution based on Division of Labor and specialization," 2022 IEEE International Conference on Internet of Things and Intelligence Systems (IoTaIS), 2022.